\documentclass[a4paper,12pt]{article}
\usepackage{amsmath,amssymb}
\usepackage{a4wide}
\usepackage{epsfig}

\def\Bbar{\smash{\overline{B}}}
\def\Kbar{\smash{\overline{K}}}
\def\BR{{\rm BR}}

\title{Nonfactorizable contributions in $\Bbar^0\rightarrow
D_s^+D_s^-$ and $\Bbar_s^0\rightarrow D^+D^-$ decays%
\footnote{Contributed Paper for LPO3}}

\author{J.O.\ Eeg$^{a}$,  S.\ Fajfer$^{b,c}$, and A.\ Hiorth$^a$\\
\footnotesize\it $^a$ Department of Physics, University of Oslo, P.O.\
Box 1048 Blindern, N-0316 Oslo, Norway\\
\footnotesize\it $^b$ \it J.\ Stefan Institute, Jamova 39, P.O.\ Box
3000, 1001 Ljubljana, Slovenia\\
\footnotesize\it $^c$ \it Department of Physics, University of
Ljubljana, Jadranska 19, 1000 Ljubljana, Slovenia}
\date{}

\begin{document}

\maketitle

\begin{abstract}
The decay amplitudes for $\Bbar^0\rightarrow D_s^+ D_s^-$ and
$\Bbar_s^0\rightarrow D^+ D^-$ have no factorizable contributions. We
suggest that dominant contributions to the decay amplitudes arise from
two chiral loop contributions and one soft gluon emission
contribution. Then we determine branching ratios $\BR(\Bbar^0 \to
D_s^+ D_s^-)\simeq 7 \times 10^{-5}$ and $\BR (\Bbar^0_s
\to D^+ D^-)\simeq 1\times 10^{-3}$.
\end{abstract}

Numerous experimental data coming from Ba Bar, Belle and Tevatron on B
meson decays stimulate many studies of their decay mechanism. Through
decades the factorization assumption has been used in calculations of
the decay amplitudes. Recently, it has been shown \cite{BBNS} that
some classes of $B$-meson decay amplitudes exhibit {\it QCD
factorization}.  This means that, up to $\alpha_s/\pi$ (calculable),
and $\Lambda_{QCD}/m_b$ (not calculable), their amplitudes factorize
into the product of two matrix elements of weak currents.  Typically,
the decay amplitudes which factorize in this sense are $B \rightarrow
\pi \pi$ and $B \rightarrow K \pi$ where the energy release is big
compared to the light meson masses.  However, for decays where the
energy release is of order 1 GeV, QCD factorization is not expected to
hold.  Here we discuss the dominant contributions in $\Bbar^0
\rightarrow D_s^+ D_s^-$ and $\Bbar_s^0 \rightarrow D^+ D^-$
\cite{EFH}.  At quark level these decays occur through the
annihilation mechanism $b \bar{s} \rightarrow c \bar{c}$ and $b
\bar{d} \rightarrow c \bar{c}$, respectively (Fig.1). However, within
the factorized limit the annihilation mechanism will give a zero
amplitude due to current conservation, as for the $D^0 \rightarrow K^0
\Kbar^0$ decay \cite{EFZ}.  The axial part of the weak current
might lead to non-zero factorized contributions if one of $D$-mesons
in the final state is a vector meson $D^*$.  Such contributions are
proportional to the numerically small Wilson coefficient $C_1$, which
we will neglect in our analysis.  In contrast, the typical factorized
decay modes which proceed through the spectator mechanism, say
$\Bbar^0 \rightarrow D^+ D_s^-$, are proportional to the
numerically larger Wilson coefficient $C_2$. If one or both of charm
mesons in this decay are vector mesons, such amplitudes will give
nonfactorizable chiral loop contributions to the process
$\Bbar_d^0\rightarrow D_s^+ D_s^-$ due to $K^0$-exchange. We
determine these chiral loop contributions.

There are also nonfactorizable contributions due to soft gluon
emission.  Such contributions can be calculated in terms of the
(lowest dimension) gluon condensate within a recently developed Heavy
Light Chiral Quark Model (HL$\chi$QM) \cite{ahjoe}, which is based on
Heavy Quark Effective Theory (HQEFT) \cite{neu}.  This model has been
applied to processes with $B$-mesons in \cite{EHP,ahjoeB}. The gluon
condensate contributions is also proportional to the favorable Wilson
coefficient $C_2$.  We follow the standard approach \cite{EffHam} for
non-leptonic decays where one constructs an effective Lagrangian
${\mathcal L}_{W}$ in terms of quark operators multiplied with Wilson
coefficients containing all information of the short distance (SD)
loop effects above a renormalization scale $\mu$ of order $m_b$.
Within Heavy Quark Effective Theory~(HQEFT)~\cite{neu}, the effective
Lagrangian ${\mathcal L}_{W}$ can be evolved down to the scale $\mu
\sim \Lambda_\chi \sim$1 GeV
\cite{GKMW,Fleischer}.

The use of factorization is illustrated in the $\Bbar^0 \rightarrow D^+ D_s^-$ decay:
\begin{eqnarray}
\langle D_s^- D^+ | {\mathcal L}_W| \Bbar^0 \rangle_F \,  = \,
 - (C_2 +\frac{1}{N_c} C_1) 
\langle D_s^- | \bar{s}\gamma_\mu \gamma_5 c |0 \rangle
\langle D^+|\bar{c}\gamma_\mu b|\Bbar^0 \rangle \; ,
\label{FactorizedP} 
\end{eqnarray}
The coefficients $C_{1,2}$ are Wilson coefficients for the operators
containing the product of two left-handed currents.  In our notation
$C_i = - \frac{G_F}{\sqrt{2}} V_{cb}V_{cs}^* a_i$, where the $a_i$ are
dimensionless, $G_F$ is the Fermi coupling, $V_{ij}$ are CKM
parameters.  Numerically, $a_1 \sim 10^{-1}$ and $a_2 \sim 1$ at the
scale $\mu = m_b$, and $|a_1| \simeq 0.4$ and $|a_2| \simeq 1.4$ at
$\mu \sim \Lambda_\chi \sim$1 GeV \cite{GKMW,Fleischer}.  Penguin
operators may also contribute, but have rather small Wilson
coefficients.

The factorized  amplitude for  $\Bbar^0\rightarrow D_s^+ D_s^-$ 
 is presented in Fig. 1, and is given by
 \begin{eqnarray}
\langle D_s^- D_s^+ | {\mathcal L}_W| \Bbar^0 \rangle_F \, = \, 
  4 (C_1 +\frac{1}{N_c} C_2) 
\langle D_s^- D_s^+|\overline{c_L} \gamma_\mu  c_L |0 \rangle
 \langle 0 |\overline{d_L} \gamma^\mu  b_L |\Bbar^0 \rangle \; .
\label{FactorizAnn}
\end{eqnarray}
Unless one or both of the $D$-mesons in the final state 
are vector mesons, this matrix
element is zero due to current conservation:
\begin{eqnarray}
\langle D_s^+ D_s^-|\overline{c} \gamma_\mu  c |0 \rangle
 \langle 0 |\overline{d} \gamma^\mu \gamma_5 b |\Bbar^0 \rangle 
 \sim f_B (p_D + p_{\bar{D}})^\mu \,
\langle D_s^+ D_s^- |\overline{c} \gamma_\mu  c |0 \rangle \; = 0 \; .
\end{eqnarray}

\begin{figure}[t]
\begin{center}
   \epsfig{file=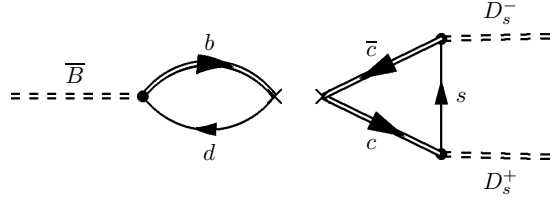,width=8cm}
\caption{Factorized contribution for 
$\Bbar^0 \rightarrow D_s^+ D_s^-$ through the annihilation mechanism,
which give zero contributions if both $D_s^+$ and $D_s^-$ are
pseudoscalars.  The double dashed lines represent heavy mesons, the
double lines represent heavy quarks, and the single lines light
quarks.}
\label{fig:bdd_fact2}
\end{center}
\end{figure}
\begin{figure}[t]
\begin{center}
   \epsfig{file=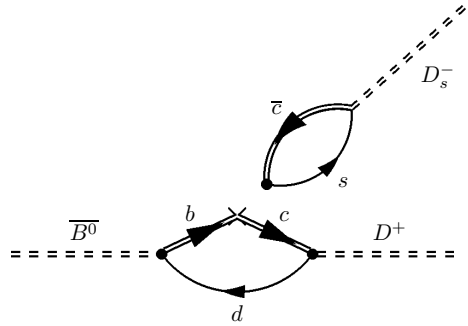,width=7cm}
\caption{Factorized contribution for 
$\Bbar^0  \rightarrow D^+ D_s^-$
through the spectator mechanism, which does not exist for
 decay mode $\Bbar^0 \rightarrow D_s^+ D_s^-$
we consider in this paper.}
\label{fig:bdd_fact}
\end{center}
\end{figure}
Our approach is based on the  use of  bosinized currents 
\cite{EFH} and by using them we first write dawn the amplitude for 
$\Bbar^0 \rightarrow D^+ D_s^-$. 
To calculate the chiral loop amplitudes
 we need the factorized amplitudes for 
$\Bbar_s^{*0} \rightarrow D_s^+ D^{*-}$ and
$\Bbar^0 \rightarrow D^{*+} D^{*-}$, which proceed through the
spectator mechanism as in  Fig.~\ref{fig:bdd_fact}.
In this case the leading chiral coupling results from 
the coupling between a pseudoscalar meson $H$, vector meson $H^*$ 
a light pseudoscalar $M$ ($=\pi, K, \eta$), denoted by $g_A$.
After use of bosonized currents \cite{EFH}, we obtain the following chiral loop amplitude for the process
 $\Bbar^0 \rightarrow D_s^+ D_s^-$ 
from the Fig. 3:
\begin{eqnarray}
A(\overline{B^0} \rightarrow D_s^+ D_s^-)_\chi \, = \,  
\left(V_{cd}^*/V_{cs}^*\right)
 \, A(\overline{B_d^0} \rightarrow D_d^+ D_s^-)_F \cdot
\; R^\chi \; \, ,
\label{chiralR}
\end{eqnarray}
where the   $A(\Bbar_d^0 \rightarrow D_d^+ D_s^-)_F$ 
stands for the  factorized amplitude
for the process $\Bbar^0 \rightarrow D^+ D_s^-$ 
 and the quantity $R^\chi$ is a sum of contributions 
from the left and right
part of Fig. 3  respectively \cite{EFH}.
In the $\overline{MS}$ scheme we obtained 
\begin{eqnarray}
R^\chi \; = \;
&&\frac{m_K^2}{(4\pi f)^2}g_A^2\left[\left\{
\frac{(\omega+1)}{(\omega+\lambda)}\, [r(-\omega)+r(-\lambda)]
- 1\right\}
\ln\left(\frac{m_K^2}{\Lambda_\chi^2}\right)- 1 \right] \; .
\label{chiralT}
\end{eqnarray}
with $\omega= M_B/(2M_D)$,  
 $\lambda=[M_B^2/(2M_D^2) -1]$ and $f$ being the $\pi$ decay constant.
The function $r(x)$ is :
\begin{equation}
r(x) \equiv \frac{1}{\sqrt{x^2-1}}\, 
\text{ln}\left(x+\sqrt{x^2-1}\right) \quad , \; \; \; 
r(-x)= -r(x) + \frac{i \pi}{\sqrt{x^2-1}} \, \; ,
\end{equation}
which  means that the amplitude
 gets an imaginary part. 
Numerically, we find \cite{EFH}:
\begin{eqnarray}
R^\chi \;\simeq \; 0.12 - 0.26 i  \; .
\label{XNum}
\end{eqnarray}
\begin{figure}[t]
\begin{center}
   \epsfig{file=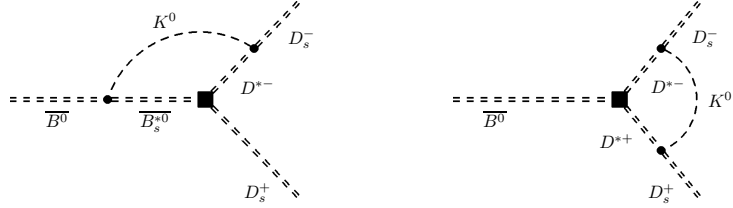,width=10cm}
\caption{Nonfactorizable chiral loops for 
$\Bbar^0 \rightarrow D_s^+ D_s^-$.}
\label{fig:chiral1}
\end{center}
\end{figure}
The genuine nonfactorizable part for 
 $\Bbar^0 \rightarrow D_s^+ D_s^-$
can, by means of Fierz transformations and identities for the product
 of two color matrices,  be written in terms of  colored currents
\begin{eqnarray}
\langle D_s^- D_s^+ | {\mathcal L}_W| \Bbar^0 \rangle_{NF} \, =  \,
   8 \, C_2 \, 
\langle D_s^- D_s^+ | (\overline{d}_L \gamma^\alpha t^a  b_L )  \; 
       ( \overline{c}_L \gamma_\alpha t^a c_L ) \  | \Bbar^0 \rangle
\; .
\label{QNFact}
\end{eqnarray}
  Within our approach, this amplitude  is written in a
 quasi-factorized way
in terms of matrix elements of colored currents:
\begin{eqnarray}
\langle D_s^+ D_s^- | {\mathcal L}_W| \Bbar^0 \rangle_{NF}^G \, = \,
  8 \, C_2 \, 
\langle D_s^+ D_s^-|\overline{c}_L \gamma_\mu t^a c_L |G \rangle
 \langle G |\overline{d}_L \gamma^\mu  t^a b_L |\Bbar^0 \rangle
\; ,
\label{QGlue}
\end{eqnarray}
where a $G$ in the bra-kets symbolizes emission of one gluon 
(from each current) as visualized in Fig. 4. 
\begin{figure}[t]
\begin{center}
   \epsfig{file=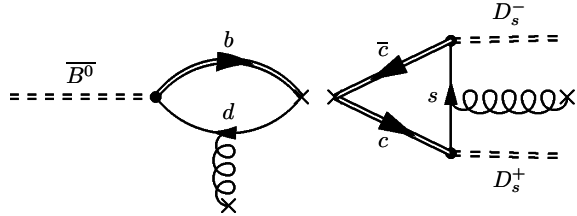,width=8cm}
\caption{Nonfactorisable contribution for 
$\Bbar^0 \rightarrow D_s^+ D_s^-$
through the annihilation mechanism with additional soft gluon emission.
 The wavy lines represent soft
gluons ending in vacuum to make gluon condensates.}
\label{fig:bdd_nfact2}
\end{center}
\end{figure}
In order to calculate the matrix elements in (\ref{QGlue}), we 
have  used  \cite{EFH} 
the Heavy Light
Chiral Quark Model (HL$\chi$QM) recently developed in \cite{ahjoe},
 which incorporates emission of soft gluons modeled by a gluon
condensate. 
Then we defined a quantity $R^G$ for the gluon
condensate amplitude  analogously to  $R^\chi$ in (\ref{chiralR})
 and (\ref{chiralT})
for chiral loops.
Numerically, we determine \cite{EFH} that the ratio between the two amplitudes
  is 
\begin{eqnarray}
R_G \; \simeq \; 0.055  + 0.16 i  \; ,
\label{RNum}
\end{eqnarray}
which is of order one third of the chiral loop contribution in
eq. (\ref{chiralT}).

Adding the amplitudes $R_\chi$ and $R_G$ and multiplying with the
Wilson coefficient \cite{GKMW,Fleischer} $a_2 \simeq 1.33 + 0.2 i$, 
we obtain the quantity:
\begin{eqnarray}
\label{Tres}
\widetilde{R_T} \, \equiv \, 
a_2 \, (R_\chi + R_G) \; \simeq 0.26 - 0.11 i \; .
\end{eqnarray}
We have found that the amplitude for
$\Bbar^0 \rightarrow D_s^+ D_s^-$ is of order $15-20 \%$
of the factorizable amplitude for
$\Bbar^0 \rightarrow D^+ D_s^-$, before the different CKM-factors are
taken into account. Finally,  we predict \cite{EFH} that 
the branching ratios are 
\begin{eqnarray}
\label{BR}
 \BR(\Bbar^0_d \to D_s^+ D_s^-) \simeq 
 7 \times 10^{-5}  \; \; ;
 \quad
 \BR(\Bbar^0_s \to D^+ D^-) \simeq 1 \times 10^{-3} \; \; .
\end{eqnarray}

The current searches at BaBar and Belle might soon result in 
the limit on the rate $\Bbar^0 \to D_s^+ D_s^-$. However,
the  $\Bbar^0_s \to D^+ D^-$ mode  will  be
 accessable at Tevatron and later at LHC.

 \vspace{0.6cm}
The research  of S.F. was supported in part
 by the Ministry of Education,
Science and Sport of the Republic of Slovenia.
J.O.E. is
 supported in part by the Norwegian research council and  by
 the European Union RTN
network, Contract No. HPRN-CT-2002-00311  (EURIDICE).


\end{document}